\documentclass[a4paper]{jpconf}
\usepackage{graphicx}
\usepackage{amsmath} 
\usepackage{amsfonts} 
\usepackage{amssymb}
\usepackage{bbm}
\newcommand{\be}{\begin{equation}}
\newcommand{\ee}{\end{equation}}
\newcommand{\ba}{\begin{eqnarray}}
\newcommand{\ea}{\end{eqnarray}}
\newcommand{\nn}{\nonumber}
\newcommand{\kr}{\rangle}
\newcommand{\kl}{\langle}

\newcommand{\cD}{{\cal D}}

\newcommand{\cN}{{\cal N}}
\newcommand{\cA}{{\cal A}}
\newcommand{\cK}{{\cal K}}

\newcommand{\cG}{{\cal G}}
\newcommand{\cR}{{\cal R}}

\begin{document}
\title{Quantum fermions and quantum field theory from classical statistics}

\author{Christof Wetterich}

\address{Institut  f\"ur Theoretische Physik\\
Universit\"at Heidelberg\\
Philosophenweg 16, D-69120 Heidelberg}

\ead{c.wetterich@thphys.uni-heidelberg.de}

\begin{abstract}
An Ising-type classical statistical ensemble can describe the quantum physics of fermions if one chooses a particular law for the time evolution of the probability distribution. It accounts for the time evolution of a quantum field theory for Dirac particles in an external electromagnetic field. This yields in the non-relativistic one-particle limit the Schr\"odinger equation for a quantum particle in a potential. Interference or tunneling arise from classical probabilities. 
\end{abstract}

\section{Introduction}
\label{Introduction}

It is often believed that quantum physics and classical statistical physics are fundamentally different on a conceptual level. For both branches of physics probabilities play a central role. Classical statistical physics involves only the concept of a positive definite probability distribution $\{p_\tau(t)\}$, rules for the computation of expectation values of observables and correlation functions in terms of $\{p_\tau(t)\}$, and a dynamical equation describing the time evolution of the probabilities. In quantum physics the complex wave function plays a central role. It is often said that quantum physics is the physics of phases in addition to the probabilities. Observables and correlations are associated to non-commuting operators. These properties lead to characteristic features as interference, entanglement, tunneling, fermions and bosons, that are all not familiar in macroscopic classical statistical physics. 

We present here a microscopic classical statistical model that accounts for all characteristic features of quantum mechanics. We consider a generalized Ising-type model for discrete Ising spins on a lattice. The discrete variables can be associated with simple yes/no alternatives or bits. The states $\tau$ of the classical statistical ensemble are configurations of Ising spins or lattice-ordered sequences of bits. The probabilities $p_\tau$ are positive and normalized, $p_\tau\geq 0$, $\Sigma_\tau p_\tau=1$. We will propose a particular evolution law for the time dependence of the probability distribution $\{p_\tau(t)\}$ that respects these properties. We will show that for this law the Ising-type classical statistical model describes a quantum field theory for Dirac-fermions in an external electromagnetic field. Specializing to probability distributions that describe one-particle states we recover the Dirac equation in an external electromagnetic field. In the limit of a static electric potential and for a non-relativistic particle this yields the Schr\"odinger equation for the quantum wave function in a potential. The characteristic features of quantum mechanics as interference in a double slit experiment are therefore accounted for by a suitable evolution law for classical probabilities. Quantum mechanics, quantum fermions and quantum field theory can be obtained from classical statistics.

This result has far reaching conceptual, and in the future perhaps also practical, consequences. Quantum systems turn out to be a particular class of classical statistical systems, characterized by particular properties of the evolution law. They are now embedded in a wider class of classical statistical systems with different evolution laws. This opens the possibility to describe phenomena as decoherence \cite{DEC} or syncoherence \cite{GR} not only by embedding a quantum subsystem into a larger quantum system that includes the environment. Many characteristic features of decoherence and syncoherence may find a description in terms of probabilities for the subsystem alone, but with an effective evolution law different from the unitary quantum evolution. Precision tests of quantum mechanics can be described in a coherent conceptual framework for ``zwitter''-particles whose properties interpolate continuously between a quantum particle and a classical particle \cite{CWP}. On the other hand, a unified conceptual framework can also help to understand the observed quantum features of interference and tunneling in experiments with classical objects, as droplets of a liquid \cite{DL}. Finally, one may hope that the equivalent classical statistical description of quantum systems may permit to device new calculational tools.

It may be worthwhile to specify clearly what we mean by a classical statistical system. The first ingredient is the classical probability distribution $\{p_\tau(t)\}$. The second one is the rule for the computation of expectation values of observables. Classical observables $A$ have a fixed value $A_\tau$ in every state $\tau$. These are the only possible outcomes of a measurement. The expectation value obeys $\kl A\kr=\sum_\tau A_\tau p_\tau$. Third, one needs to specify a law for the time evolution of $\{p_\tau(t)\}$. In our model the Ising spins are the fundamental entities. We do not assume the existence of some underlying law that determines the time evolution of a given spin configuration. Our concept is ``probabilistic realism'', where the description of nature is genuinely probabilistic. The basic dynamical law should therefore be formulated as an evolution law for the probabilities. Conceptually, this is similar to a formulation of classical mechanics with the Liouville equation as basic dynamical equation. (Newton's laws for the motion of planets can, of course, be recovered from the Liouville equation for the particular ensemble which describes isolated bodies.) In classical mechanics we can also start from Newton's equation and derive the Liouville equation, interpreting the probabilities as a ``lack of knowledge'' of a deterministic system. This possibility is not available in our setting since we have no evolution law for individual spin configurations. (It is not formally excluded that an evolution law for individual spin configurations exists, similar to cellular automata \cite{TH}. It will, however, be difficult to find it and we do not need it.) Probabilities do not reflect the lack of knowledge of an otherwise more complete system. They are simply the basic concept for formulating laws of nature. On a fundamental level we only require that the evolution law preserves positivity and normalization of the probabilities and that it is causal in the sense that the distribution $\{p_\tau(t+\epsilon)\}$ at time $t+\epsilon$ can be computed from the distribution at time $t$. This typically leads to a differential evolution equation. In practice, the proposed evolution law will also preserve locality in space. (No signals travel faster than light.)

Our derivation of the Schr\"odinger equation in terms of an Ising type classical statistical ensemble and the description of the outcome of a double slit experiment only involves these three basic ingredients of a classical statistical system. In particular, we consider ``interval observables'', which take the value one if a particle is present in a certain region of space, and zero if it is absent. These observables are directly formulated in terms of the discrete classical observables for the Ising spins. The ``discreteness'' of individual particles is therefore a natural outcome. The continuous aspects of the wave function will be related to the description of probabilities by continuous functions. Particle-wave duality finds a natural explanation in our probabilistic setting. We implement the classical statistical rule that only values $A_\tau$ in the spectrum of an observable are found in measurements. This distinguishes our approach from Bohm's quantum mechanics \cite{Bo} or other approaches where the wave function is a classical field, which has a continuous spectrum and is, in principle, observable. 

The fourth ingredient for a classical statistical system is a rule for the computation of a correlation function that describes a sequence of ideal measurements. It should be emphasized that such a rule is an independent postulate for any probabilistic description. As every experimentalist knows there are many ways to perform a sequence of two measurements - ``ideal'' ones and ``less ideal'' ones. Correlations describe the outcome of a sequence of two measurements. They depend on the precise way in which measurements are performed. ``Less ideal'' measurements will yield correlations which differ from ``ideal'' measurements. The dependence of the correlation function $\kl BA\kr$ on the precise way of performing a sequence of measurements for $A$ and $B$ is reflected mathematically in the many different possibilities to define a product structure between observables. A product structure assigns to two observables $A$ and $B$ a third observable $C$, namely $(B,A)\to C=B\circ A$. Translated to experiments, $C$ corresponds to the ``combined observable'' whose expectation value is the correlation function. In general, the correlation function $\kl BA\kr$ depends on the conditional probabilities $w(B_\rho|A_\tau)$ to find for $B$ the value $B_\rho$ if a measurement of $A$ has found the value $A_\tau$. These conditional probabilities depend on the precise way how the sequence of measurements is performed. 

Since there are many ways to perform sequences of measurements, and many associated correlation functions, one wants to define the notion of ``ideal measurements''. Correlations differing from the ones for ideal measurements can then be attributed to ``imperfect measurements''. It is at this point that some thought is needed. For macroscopic classical statistical systems an idealized sequence of measurements is generally associated to the classical correlation function $\kl BA\kr=\sum_\tau p_\tau B_\tau A_\tau$. While well adapted to large macroscopic systems, as two temperature measurements at different positions in a liquid, it often fails for isolated microscopic subsystems as an isolated atom. For an ideal measurement of properties of a subsystem the correlation function has to be computable from the information available for the description of the subsystem. This requirement often conflicts with the classical correlation function \cite{GR}. Classical correlation functions often involve properties of the subsystem as well as its environment (apparatus). The appropriate correlation functions for ideal measurements differ from the classical correlation function in these cases.

We emphasize that the issue of correlations is not relevant for the description  of the time evolution of expectation values of observables. Thus our description of interference, tunneling or entanglement in terms of the time evolution of a ``classical'' probability distribution does not involve any choice of the correlation function. This choice matters, however, if we want to describe the outcome of sequences of ideal measurements. It is important for an understanding why Bell's inequalities \cite{Bell,BS} are violated for the correlations in quantum systems \cite{GR}.

The present work is based on a series of papers by the author \cite{15}, \cite{CWF}, \cite{CWP}, \cite{GR} and has substantial overlap with ref. \cite{15}.

\section{Generalized Ising model}
\label{Generalized Ising model}

We will consider an Ising-model type system for discrete Ising-spins, $s_\gamma(\vec x)=\pm 1$, with $\vec x$ points on a suitable three-dimensional lattice and $\gamma$ denoting different ``species'' of Ising-spins, $\gamma=1\dots N_s$. For $N_s=4$ our system will be equivalent to a quantum field theory for Majorana or Weyl spinors, while for $N_s=8$ we will describe Dirac spinors. The states $\tau$ are sequences or configurations of Ising-spins, $\tau=\{s_\gamma(\vec x)\}$. Instead of Ising-spins, we will actually use occupation numbers or bits $n_\gamma(\vec x)=\big(s_\gamma(\vec x)+1\big)/2$, such that the states $\tau$ describe bit sequences of numbers $n_\gamma(\vec x)=0,1~,~\tau=\big\{n_\gamma(\vec x)\big\}$. We will consider a setting with $L^3/8$ lattice points ($L$ even) such that the number of states is $2^{(N_sL^3/8)}$. The continuum limit $L\to\infty$ is taken at the end. 

The sequences or configurations of occupation numbers $\big\{n_\gamma(\vec x)\big\}$ show already a strong analogy to the basis states of a quantum theory for an arbitrary number of fermions, formulated in the occupation number basis for positions. We will exploit this analogy in order to formulate a fundamental ``evolution law'' for the time evolution of the probability distribution $p_\tau(t)$, such that our system describes a relativistic quantum field theory for fermions.

\subsection{Classical wave function}
\label{Classical wave function}

In order to define our model we need to specify an evolution law for the time dependence of the probability distribution $\{p_\tau(t)\}$. A useful concept for this purpose is the classical wave function $\{q_\tau(t)\}$ \cite{CWP}, defined by
\be\label{35A}
q_\tau(t)=s_\tau(t)\sqrt{p_\tau(t)}~,~p_\tau(t)=q^2_\tau(t)~,~s_\tau(t)=\pm 1.
\ee
This is a real function which is given by the square roots of the probabilities up to signs $s_\tau(t)$. In terms of the wave function the expectation values of classical observables obey the ``quantum rule''
\be\label{95B}
\kl A(t)\kr=\kl q(t)\hat A q(t)\kr=\sum_\tau p_\tau(t)A_\tau,
\ee
with $\hat A$ a diagonal operator $(\hat A)_{\tau\rho}=A_\tau\delta_{\tau\rho}$. The quantum rule follows directly from the classical statistical rule.  The signs $s_\tau$ do not appear in eq. \eqref{95B}. At this stage the use of a wave function seems redundant. Nevertheless, it permits the use of the quantum formalism for a description of a classical statistical system. 

Consistent evolution laws have to correspond to rotations of the vector $q_\tau(t)$, 
\be\label{35B}
q_\tau(t)=\sum_\rho R_{\tau\rho}(t,t')q_\rho(t')~,~\sum_\rho R_{\tau\rho}(t,t')R_{\sigma\rho}(t,t')=\delta_{\tau\sigma}.
\ee
The positivity of $p_\tau=q^2_\tau$ is automatic, and the normalization of the distribution remains preserved provided it was normalized for some initial time $t_{in}$, since the length of a vector or $\Sigma_\tau q^2_\tau=\Sigma_\tau p_\tau$ is preserved by rotations. We will specify a {\em linear evolution for the wave function}. In other words, we will consider rotation matrices $R_{\tau\rho}$ that are independent of the wave function. The classical wave function specifies the probability distribution uniquely. The specification of an evolution law for the wave function therefore defines the dynamics of the classical statistical system completely. 

Quantum wave functions are usually defined in a complex Hilbert space. Indeed, many characteristic quantum features are closely associated to the ``physics of phases''. In our setting we will define later a complex quantum  wave function by introducing a complex structure in the real space spanned by $\{q_\tau\}$. The $2^{(N_sL^3/8)}$ real components of the vector $\{q_\tau\}$ can then be associated to a complex vector with dimension $2^{(N_sL^3/8)-1}$. There is no conceptual difference between the classical and quantum wave function in our setting. However, more of the information contained in a complex wave function $\psi_\alpha=\psi_{R,\alpha}+i\psi_{I,\alpha}$ will now directly be associated to classical probabilities. In quantum mechanics one usually distinguishes between the probabilities $p_\alpha=|\psi_\alpha|^2$ and the phases, whereas in our approach both positive numbers $\psi^2_{R,\alpha}$ and $\psi^2_{I,\alpha}$ will be associated to classical probabilities $p_\tau$. Our description of fermions and quantum field theory will be based entirely on the real ``classical'' wave function $\{q_\tau\}$, which corresponds to the roots of the probabilities according to eq. \eqref{35A}.

The only information contained in the classical wave function which goes beyond the set of probabilities $\{p_\tau\}$ concerns the signs $\{s_\tau\}$. The choice of $\{s_\tau\}$ does not affect the expectation values of classical observables and may therefore be considered as a type of gauge choice. A linear evolution law \eqref{35B} involves, however, directly these signs and one may wonder if the evolution of the classical wave function involves physics beyond the classical statistical setting, thereby ``smuggling in'' quantum mechanics in some hidden way. This is actually not the case. For computing $q(t+\epsilon)$ from a given wave function at time $t$ according to eq. \eqref{35B}, we only have for every $\tau$ the choice between the two values $s_\tau(t)=\pm 1$. Only one of them is consistent with the normalization of the probabilities. This fixes for all $t$ the sign distribution $\{s_\tau(t)\}$ in terms of some initial distribution'' $\{s_\tau(t_0)\}$ for $t=t_0$. For a differential evolution equation it is easy to see \cite{CWP} that $s_\tau(t)$ is constant as long as $p_\tau>0$. Possible sign jumps of $s_\tau$ can occur for particular times where $p_\tau(t)=0$, and are decided by the normalization of the probability distribution. 

One may visualize this issue by a simple periodic time evolution for a two state system with $\tau=0,1$ \cite{CWES}. A periodic change between the probabilities $p_0$ and $p_1=1-p_0$ can be described by the second order evolution equation $\partial^2_t p_0=2\omega^2(1-2p_0)$, with solution $p_0=\cos^2(\omega t+\alpha)$. In terms of the wave function this becomes a first order differential equation 
\be\label{3A1}
\partial_t
\left(\begin{array}{c}q_0\\q_1\end{array}\right)=\omega
\left(\begin{array}{r}0,1\\-1,0\end{array}\right)
\left(\begin{array}{c}q_0\\q_1\end{array}\right).
\ee
The probability distribution is fixed for all $t$ by the initial values $p_0(t_0)$ and $\partial_tp_0(t_0)$ for the second order evolution equation and equivalently by $p_0(t_0),s_0(t_0)s_1(t_0)$ for eq. \eqref{3A1}. Only the relative sign $s_0s_1$ matters, such that the needed ``initial data'' are the same for both descriptions. (It is actually possible to compute the signs $s_\tau(t)$ from ``all time probabilities'' \cite{CWES}.) For an expression of the relevant information contained in the wave function it is important that the latter is real. This contrasts with the discussion of classical mechanics in a Hilbert space by Koopman and von Neumann \cite{Kop}, where phases of a complex wave function appear as new degrees of freedom beyond the probability distribution. 

\subsection{Grassmann wave function}
\label{Grassmann wave function}

A second useful formal tool for the description of our Ising type model is a map to an equivalent description in terms of Grassmann variables. This will make the close connection to fermions most transparent \cite{CWF}. The Grassmann formulation relies on the isomorphism between states $\tau$ and the basis elements $g_\tau$ of a real Grassmann algebra that can be constructed from the Grassmann variables $\psi_\gamma(x)$. Each basis element $g_\tau$ is a product of Grassmann variables $g_\tau=\psi_{\gamma_1}(x_1)\psi_{\gamma_2}(x_2)\dots$ which is ordered in some convenient way. If a Grassmann element $g_\tau$ contains a given variable $\psi_\gamma(x)$ we put the number $n_\gamma(x)$ in the sequence $\tau$ to $0$, while we take $n_\gamma(x)=1$ if the variable $\psi_\gamma(x)$ does not appear in the product. 

An arbitrary element $g$ of the Grassmann algebra can be expanded in terms of the basis elements 
\be\label{YA}
g=\sum_\tau q_\tau g_\tau.
\ee
A time dependent wave function $\big\{q_\tau(t)\big\}$ can therefore be associated to a time dependent element of the Grassmann algebra $g(t)$, provided the coefficients $q_\tau(t)$ are real and normalized according to $\sum_\tau q^2_\tau(t)=1$. We will formulate the fundamental evolution law as an evolution law for the ``Grassmann wave function'' $g(t)$. 

For this purpose we will formulate in the next section a Grassmann functional integral for a quantum field theory of massless Majorana fermions. It will involve $N_sL^3/8$ Grassmann variables $\psi_\gamma(t,x)$ for every discrete time point $t$. The Grassmann wave function $g(t)$ and the fundamental evolution law can be extracted from this functional integral. This will establish a map between a quantum field theory for fermions and Ising type classical statistical ensembles with dynamics specified by an appropriate evolution law. From the point of view of the Ising type classical statistical ensemble the functional integral is a pure technical device for the specification of an evolution law for the classical wave function $\{q_\tau(t)\}$ and probability distribution $\{p_\tau(t)\}$. Conceptually, we could start equivalently by specifying the matrix $R_{\tau\rho}(t,t-\epsilon)$ in eq. \eqref{35B}. In practice, the functional integral formulation is much simpler since it provides for an easy realization of appropriate symmetries of the evolution law, as Lorentz symmetry in our particular model.

\section{Evolution law, action and functional integral}
\label{Evolution law, action and functional integral}

\subsection{Action and functional integral}
\label{Action and functional integral}

We start with the action $S$ and Lagrangian $L(t)$ for the regularized quantum field theory of massless Majorana spinors $(N_s=4)$
\be\label{N1}
S=\sum^{t_f-\epsilon}_{t=t_{in}}L(t)~,~L(t)=\sum_x\psi_\gamma(t,x)B_\gamma(t+\epsilon,x).
\ee
Here we define
\be\label{7A}
B_\gamma(t+\epsilon,x)=\prod_j\Big(\sum_{v_j=\pm 1}\Big)
Y_{\gamma\delta}\big(\{v\}\big)\psi_\delta(t+\epsilon,x_k+v_k\Delta),
\ee
with
\be\label{B17}
Y_{\gamma\delta}\big(\{v\})=\frac18
\Big[1-\sum_k(v_k+w_k\tilde I)T_k-v_1v_2v_3\tilde I\Big]_{\gamma\delta}.
\ee
Thus $B_\gamma(t+\epsilon,x)$ involves a linear combination of Grassmann variables at lattice sites which are diagonal neighbors of $x$, corresponding to corners of a cube with basis length $2\Delta$, with $x$ being at its center. The sums extend over $j,k=1\dots 3$ and each corner corresponds to a particular combination of the three signs $v_j=\pm 1$. We define $w_k$ by $w_1=-v_2v_3,w_2=v_1v_3$, $w_3=-v_1v_2$. Eqs. \eqref{N1}-\eqref{B17} and the following imply a summation over repeated indices $\gamma,\delta=1\dots 4$. 

We have also introduced the three real symmetric $4\times 4$ matrices $T_k$ as 
\be\label{18AA}
T_1={0,1\choose 1,0}~,~T_2={~~0,c\choose -c,0}~,~
T_3={1,~~0\choose 0,-1}~,~T^T_k=T_k,
\ee
where $c=i\tau_2,\tau_k$ are Pauli matrices and $1$ stands for the unit $2\times 2$ matrix. The product of these matrices yields the real antisymmetric $4\times 4$ matrix 
\be\label{18ABa}
\tilde I=-{c,0\choose~ 0,c}=T_1T_2T_3~,~\tilde I^T=-\tilde I.
\ee

The sum in eq. \eqref{N1} extends over discrete time points $t_n$, with $\int_t=\epsilon\sum_t=\epsilon\sum_n~,~t_{n+1}-t_n=\epsilon,n\in{\mathbbm Z}$, $t_{in}\leq t_n\leq t_f$. The time-continuum limit is taken as $\epsilon\to 0$ for fixed $t_{in},t_f$. Similarly, we sum in eq. \eqref{N1} over points $x$ of a lattice. For this purpose we consider for every given $t$ a cubic lattice with lattice distance $2\Delta$ and $\int_x=8\Delta^3\sum_x$. We may take $\epsilon =\Delta$ such that the space-time lattice points belong to a hypercubic bcc lattice with distance $2\Delta$ between nearest neighbors, which we call the ``fundamental lattice''. For even $t=2n\epsilon$ the space lattice points are even, $x_k=2m'_k\Delta$, with $n,m'_k\in{\mathbbm Z}$. This will be called the even sublattice. The odd sublattice consists of the odd time points $t=(2n+1)\epsilon$ for which the space points are also odd, $x_k=(2m'_k+1)\Delta$.  The action \eqref{N1} involves indeed only Grassmann variables for points on the fundamental lattice. For even $t$ eq. \eqref{N1} involves Grassmann variables $\psi_\gamma(t,x)$ living on the even sublattice, while the combination $B_\gamma(t+\epsilon,x)$ lives on the odd sublattice. For odd $t$ the role of the sublattices is exchanged, now with $\psi$ on the odd and $B$ on the even sublattice. This construction eliminates ``lattice doublers'' - a more detailed discussion of the lattice implementation can be found in ref. \cite{15}.

The action \eqref{N1} is an element of a real Grassmann algebra - all coefficients multiplying the Grassmann variables $\psi_\gamma(t,x)$ are real. Within the Grassmann algebra the operation of transposition amounts to a total reordering of all Grassmann variables. The action \eqref{N1} is antisymmetric under this operation, $S^T=-S$. If we define formally the Minkowski action $S_M=iS$ the latter is hermitean, $S_M=S^\dagger_M$ since $S^*_M=-S_M$. 

The action \eqref{N1} is Lorentz-symmetric in the continuum limit $\epsilon=\Delta,\Delta\to 0$. Indeed, performing the continuum limit the action \eqref{N1}-\eqref{B17} becomes \cite{15}
\be\label{F1}
S=\int_{t,x}\big\{\psi_\gamma\partial_t\psi_\gamma-\psi_\gamma
(T_k)_{\gamma\delta}
\partial_k\psi_\delta\big\}.
\ee
It involves now four Grassmann functions $\psi_\gamma(t,x),\gamma=1\dots 4,x=(x_1,x_2,x_3)$. The integral extends over three dimensional space and time, with $\partial_t=\partial/\partial t$ and $\partial_k=\partial/\partial x_k$. The Lorentz invariance of the action \eqref{F1} is most easily established by employing the real matrices
\be\label{F7}
\gamma^0=\left(\begin{array}{rcr}
0&,&\tau_1\\-\tau_1&,&0\end{array}\right)~,~
\gamma^k=-\gamma^0 T_k,
\ee
such that
\be\label{F8}
S=-\int_{t,x}\bar\psi\gamma^\mu \partial_\mu \psi~,~\bar\psi=\psi^T\gamma^0,
\ee
where $\mu=(0,k)$ and $\partial_0=\partial_t$. The real $4\times 4$ Dirac matrices $\gamma^\mu$ obey the Clifford algebra, 
$\{\gamma^\mu,\gamma^\nu\}=2\eta^{\mu\nu}$, with signature of the metric given by $\eta_{\mu\nu}=diag(-1,1,1,1)$. 

The functional integral is defined by the partition function
\be\label{N9}
Z=\int{\cal D}\psi\bar g_f\big[\psi(t_f)\big]e^{-S}
g_{in}\big[\psi(t_{in})\big]~,~
\int{\cal D}\psi=\prod_{t,x}\int \big(d\psi_4(t,x)\dots d\psi_1(t,x)\big).
\ee
The boundary terms $g_{in}$ and $\bar g_f$ only depend on the Grassmann variables $\psi(t_{in})$ and $\psi(t_f)$, respectively. The boundary terms $\bar g_f$ and $g_{in}$ are related to each other \cite{15}, such that the functional integral \eqref{N9} is fully specified by the choice of $g_{in}$. In turn, $g_{in}$ is given by the initial values for the classical wave function, cf. eq. \eqref{YA}.

\subsection{Grassmann wave function from functional integral}
\label{Grassmann wave function}

In this section we compute for the functional integral \eqref{N9} a Grassmann wave function $g(t)$, which is an element of the Grassmann algebra constructed from the Grassmann variables $\psi_\gamma(t,x)$ at given $t$. The functional integral \eqref{N9} involves variables for arbitrary time points $t_n$. In order to construct a wave function $g(t)$ which only refers to a particular time $t$ we have to ``integrate out'' the information referring to other time points $t'\neq t$ \cite{CWF,14}. This can be done by decomposing the action \eqref{N1}
\be\label{M1}
S=S_<+S_>~,~S_<=\sum_{t'<t}L(t')~,~S_>=\sum_{t'\geq t}L(t').
\ee
The wave function $g(t)$ obtains now by integrating out all Grassmann variables for $t'<t$
\be\label{M2}
g(t)=\int {\cal D}\psi(t'<t)e^{-S_<}g_{in}.
\ee

As a consequence of its definition \eqref{M2} the Grassmann wave function obeys the time evolution
\ba\label{P1}
g(t+\epsilon)=\int\cD\psi(t)e^{-L(t)}g(t)=\int\cD\psi(t)\exp\big\{-\sum_x\sum_\gamma\psi_\gamma(t,x)
B_\gamma(t+\epsilon,x)\big\}g(t).
\ea
This determines $\big\{q_\tau(t+\epsilon)\big\}$ in terms of $\big\{q_\tau(t)\big\}$. Thus the action \eqref{N1}-\eqref{B17} specifies the dynamics how the probability distribution $\big\{p_\tau(t)\big\}$ evolves in time. The particular dynamics of a given model is determined by the form of $B_\gamma$ in eq. \eqref{B17}. In a similar way we can relate $g(t+2\epsilon)$ to $g(t)$. Performing the necessary Grassmann integrals over $\psi(t)$ and $\psi(t+\epsilon)$ one obtains \cite{15} for $\epsilon\to 0$ the differential evolution equation
\be\label{P34}
\partial_tg(t)=\frac{1}{2\epsilon}\big[g(t+2\epsilon)-g(t)\big]={\cal K} g(t)=\sum_\tau q_\tau (t){\cal K} g_\tau 
=\sum_{\tau,\rho}q_\tau(t)g_\rho
K_{\rho\tau}=\sum_\tau\partial_t q_\tau(t)g_\tau, 
\ee
with Grassmann evolution generator ${\cal K}$
\be\label{P38}
{\cal K}=\sum_x\frac{\partial}{\partial\psi_\gamma(x)}(T_k)_{\gamma\delta}
\partial_k\psi_\delta(x).
\ee
From eq. \eqref{P34} we can directly extract the evolution equation for the classical wave function \eqref{35B} in a continuous version,
\be\label{P35}
\partial_t q_\tau(t)=\sum_\rho K_{\tau\rho} q_\rho(t)~,~K_{\rho\tau}=\int\cD\psi\tilde g_\rho{\cal K} g_\tau.
\ee
We identify this evolution equation with the generalized Schr\"odinger equation for a quantum wave function for the special case of a real wave function and purely imaginary and hermitean Hamiltonian $H=i\hbar K$. It specifies the evolution of the probabilities for the Ising-type classical statistical ensemble. 

\section{Majorana fermions}
\label{Majorana fermions}

\subsection{Conserved quantities}
\label{Conserved quantities}

For a physical interpretation of the evolution equation \eqref{P34}, \eqref{P38} it is useful to look for conserved quantities. They correspond to time independent expectation values of classical statistical observables. We first represent the classical observables in the Grassmann formalism. Consider the observable measuring the occupation number $N_\gamma(x)$ of the bit $\gamma$ located at $x$. The spectrum  of possible outcomes of measurements consists of values $1$ or $0$, depending if a given state $\tau=[n_\gamma(x)]$ has this particular bit occupied or empty. For the Grassmann basis element $g_\tau$ associated to $\tau$ one finds $N_\gamma(x)=0$ if $g_\tau$ contains a factor $\psi_\gamma(x)$, and $N_\gamma(x)=1$ otherwise. We can associate to this observable a Grassmann operator ${\cal N}_\gamma(x)$ obeying (no summation here) 
\be\label{G16}
\cN_\gamma(x)g_\tau=\big(N_\gamma(x)\big)_\tau g_\tau~,~\cN_\gamma(x)=
\frac{\partial}{\partial\psi_\gamma(x)}
\psi_\gamma(x).
\ee
Two occupation number operators ${\cal N}_{\gamma_1}(x_1)$ and ${\cal N}_{\gamma_2}(x_2)$ commute. 

In general, we may associate to each classical observable $A$ a diagonal quantum operator $\hat A$ acting on the wave function, defined by $(\hat A q)_\tau=A_\tau q_\tau$ with $\hat A$ a diagonal operator $\hat A_{\tau\rho}=A_\tau\delta_{\tau\rho}$. In the Grassmann formulation one uses the associated Grassmann operator ${\cal A}$ obeying
\be\label{G18}
\cA g_\tau=A_\tau g_\tau~,~\kl A\kr=\int \cD\psi\tilde g\cA g.
\ee
Here $\tilde g$ is conjugate to $g$, i.e. for $g=\sum_\tau q_\tau g_\tau$ one has $\tilde g=\sum_\tau q_\tau\tilde g_\tau$. The conjugate basis elements of the Grassmann algebra $\tilde g_\tau$ are defined \cite{CWF} by the relation
\be\label{M6}
\tilde g_\tau g_\tau=|0\kr=\prod_x\prod_\gamma\psi_\gamma(x)~,~\int {\cal D}\psi|0\kr=1,
\ee
(no sum over $\tau$) and the requirement that no variable $\psi_\gamma(x)$ appears both in $\tilde g_\tau$ and $g_\tau$. 

Using $\partial_t q=Kq$ \eqref{P35} and $\partial_t g={\cal K} g$ \eqref{P34} the time evolution of the expectation value is expressed in terms of the commutators
\be\label{95D}
\partial_t\kl A\kr=\kl q[\hat A,K]q\kr=\int\cD\psi\tilde g[{\cal A},{\cal K}]g.
\ee
Conserved quantities are represented by Grassmann operators that commute with ${\cal K},[{\cal A},{\cal K}]=0.$

\subsection{Particle states}
\label{Particlestates}

Our system admits a conserved particle number, corresponding to the sum over all occupation numbers $N_\gamma(x)$ in the lattice, or to the Grassmann operator $\cN$,
\be\label{H5}
\cN=\int_y\frac{\partial}{\partial\psi_\gamma(y)}\psi_\gamma(y)~,~[\cN,\cK]=0.
\ee
We can decompose an arbitrary Grassmann element into eigenstates of $\cN$, $g=\Sigma_mA_mg_m~,~\cN g_m=mg_m$. For an eigenstate $g_m$ the probabilities in the Ising-type classical statistical ensemble differ from zero only if the total number of ``occupied bits'' equals precisely $m$. The time evolution does not mix sectors with different particle number $m$, such that the coefficients $A_m$ are time independent, $\partial_t g=\Sigma_mA_m\partial_tg_m~,~\partial_tg_m=\cK g_m$. We can restrict our discussion to eigenstates of $\cN$.

A static vacuum state $g_0$ with a fixed particle number $m_0$ obeys
\be\label{H8}
\cK g_0=0~,~\cN g_0=m_0g_0~,~\int \cD\psi \tilde g_0 g_0=1.
\ee
An example for a possible vacuum state is the totally empty state $g_0=|0\kr$ (eq. \eqref{M6}). Another example is the totally occupied state with $g_0=1,~m_0=B=N_s L^3/8$. There are many possible vacuum states, for example a half filled state with $m_0=B/2$. 

We next define creation and and annihilation operators $a^\dagger_\gamma(x),~a_\gamma(x)$ as
\be\label{47A}
a^\dagger_\gamma(x)g=\frac{\partial}{\partial\psi_\gamma(x)}g~,~a_\gamma(x)g=
\psi_\gamma(x)g~,~
\big \{a^\dagger_\gamma(x),~a_\epsilon(y)\big\}=\delta_{\gamma\epsilon}\delta(x-y)~,~
\cN=\int_xa^\dagger_\gamma(x)a_\gamma(x).
\ee
Acting with the creation operator on the vacuum produces one-particle states
\be\label{H7a}
g_1(t)=\int_xq_\gamma(t,x)a^\dagger_\gamma(x)g_0=\cG_1g_0~,~(\cN-m_0)g_1=g_1.
\ee
The time evolution of the one-particle wave function $q_\gamma$ is given by
\be\label{H11} 
\partial_t g_1=\cK g_1=\int_x 
q\left[\cK,\frac{\partial}{\partial\psi}\right]g_0
=\int_x(\partial_tq\frac{\partial}{\partial\psi})g_0.
\ee
Inserting the specific form of the evolution operator \eqref{P38} yields the Dirac equations for a real wave function \cite{CWF}
\be\label{H13}
\gamma^\mu \partial_\mu q=0.
\ee
We emphasize that the Dirac equation follows for arbitrary static states $g_0$ which obey ${\cal K}g_0=0$. Since there is no mass term and the Dirac-matrices as well as the wave function are real, eq. \eqref{H13} describes the time evolution of the wave function for a single massless Majorana fermion.  

States with of $n$ fermions can be constructed by applying $n$ creation operators $a^\dagger_\gamma(x)$ on the vacuum. For example, the two-fermion state obeys
\be\label{H33}
g_2(t)=\frac{1}{\sqrt{2}}\int_{x,y}q_{\gamma\epsilon}(t,x,y)a^\dagger_\gamma(x)
a^\dagger_\epsilon(y)g_0.
\ee
Since the creation operators anticommute the two-particle wave function is antisymmetric, as appropriate for fermions. From there we recover the standard interference rules for identical fermions. The linearity of the evolution equation in $q$ is crucial in this respect. It guarantees the superposition principle for the wave function, and therefore interference effects for the probabilities.

\section{Dirac fermions in electromagnetic fields}
\label{Diracfermions}

Dirac spinors can be composed of two different Majorana spinors with equal mass $(N_s=8,B=L^3)$. We extend our model by introducing in the action a mass term $\sim m$ and a coupling $\sim e$ to an external electromagnetic potential $A_\mu(x)$, (with suppressed spinor indices)
\ba\label{E1}
S&=&\int_{t,x}\big\{\psi_1(\partial_t-T_k\partial_k+m\gamma^0\tilde I)\psi_1
+\psi_2(\partial_t-T_k\partial_k+m\gamma^0\tilde I)\psi_2\big\}\nn\\
&&-e\int_{t,x}\big\{\psi_1(A_0-A_kT_k)\psi_2
-\psi_2(A_0-A_kT_k)\psi_1\big\}.\label{E13}
\ea

We introduce a complex structure by defining the four-component complex Grassmann variables $\psi_D=\psi_1+i\psi_2~,~\psi^*_D=\psi_1-i\psi_2$. In terms of $\psi_D$  the action \eqref{E1}  takes the familiar form 
\ba\label{E4}
S=-\int_{t,x}\bar\psi_D(\gamma^\mu\partial_\mu-m\tilde I)\psi_D -ie\int_{t,x}\bar\psi_D\gamma^\mu A_\mu\psi_D~,~\bar\psi_D=\psi^\dagger_D\gamma^0.
\ea

\subsection{Evolution equation for Dirac fermions}  
\label{Evolution equation for Dirac fermions}

For the action \eqref{E1} the Grassmann evolution equation $\partial_t g={\cal K}g$ involves ${\cal K}={\cal K}_0+\cK_m+{\cal K}_A$
\ba\label{X1}
{\cal K}_0+\cK_m&=&\int_x\sum_{a=1,2}\frac{\partial}{\partial\psi_a(x)}
(T_k\partial_k-m\gamma^0\tilde I)\psi_a(x),\\
{\cal K}_A&=&e\int_x\left[\frac{\partial}{\partial\psi_1(x)}
(A_0(x)-A_k(x) T_k)\psi_2(x)-\frac{\partial}{\partial\psi_2(x)}
(A_0(x)-A_k(x)T_k)\psi_1(x)\right].\nn
\ea
It describes the dynamics for an arbitrary number of charged relativistic fermions (and their antiparticles) in external electromagnetic fields. 

The evolution equation for the classical wave function \eqref{P35}, and therefore for the probability distribution, obtains again from eq. \eqref{P34}. In summary, we have formulated a time evolution equation for a classical wave function $q_\tau(t)$ which describes an arbitrary number of charged electrons in external electric and magnetic fields. Its restriction to one-particle states yields the relativistic Dirac equation. As usual, a non-relativistic approximation will yield the familiar Schr\"odinger equation for a particle in a potential or moving in external magnetic fields. 

\subsection{Classical observables}
\label{Classical observables}

Classical observables can be constructed from linear combinations of products of occupation numbers. We combine the index $a=1,2$ for the two Majorana spinors $\psi_a$ with the index $\gamma$ into a common index $\epsilon=(\gamma,a),\epsilon=1\dots 8$. The Grassmann operators corresponding to the occupation numbers $N_\epsilon(x)$ obey eq. \eqref{G16}. A local particle number can be defined as 
\be\label{FF8}
\cN(x)=\sum_\epsilon\cN_\epsilon(x)~,~
\big[\cN(x),\cK \big]=(T_k)_{\eta\alpha}\partial_k{\cal M}_{\alpha\eta}~,~{\cal M}_{\epsilon\eta}(x)=\frac{\partial}{\partial\psi_\epsilon(x)}\psi_\eta(x).
\ee
According to eq. \eqref{95D} the time evolution of the classical mean local particle number obeys 
\be\label{FF10}
\partial_t\kl N(x)\kr=(T_k)_{\eta\alpha}\partial_k\kl {\cal M}_{\alpha\eta}\kr.
\ee
This involves the expectation values of ``off-diagonal'' operators ${\cal M}_{\alpha\eta}$. Products of operators ${\cal M}_{\alpha\eta}$ can be defined by the non-commutative quantum product and we see here a first appearance of non-commutative structures. The total particle number $\cN=\int_x\cN(x)~,~[\cN,\cK]=0$, is conserved.

\section{Quantum mechanics for particle in a potential}
\label{Quantum mechanics for particle in a potential}

\subsection{Dirac equation}
\label{Dirac equation}

We next concentrate on one-particle states, described by Grassmann elements
\be\label{X2}
g_1(t)=\int_x\left(q_{1,\gamma}(t,x)
\frac{\partial}{\partial\psi_{1,\gamma}(x)}+q_{2,\gamma}(t,x)
\frac{\partial}{\partial\psi_{2,\gamma}(x)}\right)g_0.
\ee
Here $g_0$ is some arbitrary static ``vacuum state''. For ${\cal K}g_0=0$ we find for the one-particle wave function the Dirac equation
\be\label{276C}
\gamma^\mu(\partial_\mu+ieA_\mu)\varphi_D=im\bar\gamma\varphi_D~,~\varphi_D=q_1+iq_2.
\ee
This equation holds for an arbitrary representation of the Dirac matrices $\gamma^\mu$. The matrix $\bar\gamma$ (often called $\gamma^5$) obeys $\bar\gamma=-i\gamma^0\gamma^1\gamma^2\gamma^3$. In our conventions \cite{CWMSab} the Dirac equation obeys parity, charge conjugation and time reversal symmetry.

The derivation of the Dirac equation for a one-particle state \eqref{X2} has only used the general evolution equation for Grassmann elements $g(t)$ and the condition $\cK g_0=0$. It therefore describes the dynamics for a very extended family of classical probability distributions or classical wave functions. Indeed, it is sufficient that $g_0$ is an arbitrary static state (not necessarily a priori with a fixed particle number). This reflects the physical property that isolated one-particle states can occur under a wide variety of circumstances, and that for sufficient isolation the properties of the environment do not matter for the dynamics of the isolated particle. In this context we emphasize, however, that our model only describes external electromagnetic fields while we do not account for the fields generated by the particles that may be present  in the environment. In this sense our setting describes ``real physics'' only in a situation where the electromagnetic fields generated by all present particles can be approximated by a ``mean field'' that is independent of individual particle positions. This is precisely the setting of standard one-particle quantum mechanics.

\subsection{Particle observables}
\label{Particle observables}

Let us consider the expectation value of the local occupation number ${\cal N}(x)$ defined by eq. \eqref{FF8}. For $g_0=|0\kr$ its expectation value obeys the simple relation 
\be\label{VV3}
\kl \cN(x)\kr=\sum_\epsilon q^2_\epsilon(t,x)~,~\int_x \sum_\epsilon\big(q_\epsilon(x)\big)^2=\int_x\varphi^\dagger_D(x)\varphi_D(x)=1.
\ee
The normalization \eqref{VV3} is preserved by the unitary time evolution of the one-particle wave function. Indeed, we can write the Dirac equation \eqref{276C} as a Schr\"odinger-type equation $i\hbar\partial_t\varphi_D=H\varphi_D$, with hermitean Hamiltonian $H=i\hbar T_k\partial_k+\hbar m\gamma^0\bar\gamma+\hbar e(A_0 -T_k A_k)$.

For a pure one-particle state the expectation value $\kl N(x)\kr$ can be interpreted in a natural way as the probability density to find the particle at the position $x$. This is precisely the standard interpretation of $\psi^\dagger_D(x)\psi_D(x)$ in one-particle quantum mechanics,
\ba\label{VV9}
w(x)=\kl N(x)\kr=\varphi^\dagger_D(x)\varphi_D(x)~,~
\int_xw(x)=1.
\ea
Since $N(x)$ is a classical observable, we can define the position of the particle as a classical observable, and similarly for functions of $X$,
\be\label{VV10}
X=\int_x xN(x)~,~f(X)=\int_xf(x)N(x).
\ee
The expectation value in classical statistics coincides with the standard quantum mechanics rule
\be\label{VV11}
\kl X\kr=\kl \int_x xN(x)\kr=\int_xxw(x)=\int_x\varphi^\dagger_D(x)x\varphi_D(x)~,~
\kl f(X)\kr=\int_x\varphi^\dagger_D(x)f(x)\varphi_D(x).
\ee
In particular, one obtains the same formula for the dispersion $\kl X_kX_k\kr-\kl X_k\kr\kl X_k\kr$ as in quantum mechanics. We conclude that measurements of the position of a particle, or more generally the distribution of positions in an ensemble of one-particle states, can be described equivalently in an Ising type a classical statistical ensemble with classical observables, or in quantum mechanics. The complete time evolution of the distribution of positions is identical in both descriptions. This covers, in particular, the characteristic quantum interference in a double slit experiment. The time evolution \eqref{P35}, \eqref{276C} for the classical wave function and associated classical probability distribution $\{p_\tau\}$ produces for one-particle states exactly the quantum mechanical interference pattern.

\subsection{Schr\"odinger equation}
\label{equation}

Standard quantum mechanics for an electron in a potential is recovered from the non-relativistic approximation to the Dirac equation. This is well known, and we sketch here for completeness only the case $A_k=0$. The nonrelativistic approximation becomes valid if $eA_0$ and $iT_k\partial_k$ are small compared to $m$. Since $(\gamma^0\bar \gamma)^2=1$, it is convenient to choose a new basis where $\gamma^0\bar\gamma=diag (1,1,-1,-1)$. With $V(x)=\hbar eA_0(x)$ and $M=\hbar m$ the Hamiltonian takes the form
\be\label{VV14} 
H=\left(\begin{array}{ccc}M&,&\sigma_kp_k\\\sigma^\dagger_kp_k&,&-M\end{array}\right)+V(x)~,~p_k=-i\hbar \partial_k,
\ee
with matrices $\sigma_1=-1~,~\sigma_2=i\tau_2~,~\sigma_3=i\tau_3$. The appearance of $\hbar$ only fixes the units for $p_k,M$ and $V(x)$.

We next decompose $\varphi_D$ into two-component wave functions, $\varphi^T_D=(\chi^T,\rho^T)$. For the non-relativistic electron we consider the approximate solution $\rho=A\chi$, where $A=\sigma^\dagger_kp_k/(2M)$ is determined by requiring in leading order $\partial_t\rho=A\partial_t\chi$. Insertion into eq.  \eqref{276C} yields for $\psi=\exp(iMt/\hbar)\chi$ the standard Schr\"odinger equation for a particle in a potential $V$,
\be\label{VV22}
i\hbar\partial_t\psi=\left(\frac{p_kp_k}{2M}+V\right)\psi.
\ee

All quantum mechanical phenomena extracted from solutions of the Schr\"odinger equation are described by our time evolution equation for a classical statistical ensemble of Ising-spins on a lattice. For a potential realizing a double-slit situation the standard interference pattern will appear behind the slits. This holds provided that the initial state at some time $t_0$ corresponds to a one-particle state describing a particle moving towards the slits. Interference is realized by classical probabilities. 

For the choice $g_0=|0\kr$ the evolution law for the classical probabilities associated to a one particle state is rather simple. The only non-vanishing probabilities $p_\epsilon (x)=q^2_\epsilon(x)$ occur if precisely one lattice site is occupied with a given species $\epsilon$. The evolution of $q_\epsilon(t,x)$ obeys eq. \eqref{276C}. We recall, however, that the same physical situation arises for arbitrary static $g_0$. The time evolution of probabilities can then get rather complicated in the Ising-spin picture whereas it remains simple in the fermionic picture. 

\subsection{Particle-wave duality}
\label{Particle-wave duality}

The discreteness of measurement values in quantum mechanics can be traced back to the discrete occupation numbers of the Ising-type model. In quantum mechanics we may define an ``interval observable'' $J_{{\cal R}}$ by a function
\be\label{VV24}
J_\cR(x)=\left\{\begin{array}{ll}
1&\text{if}~x\in \cR\\0&\text{otherwise}
\end{array}\right)~,~J^2_\cR=J_\cR,
\ee
such that its spectrum consists of the discrete values $0$ and $1$. According to the rules of quantum mechanics, the possible outcomes of a measurement of $J_\cR$ are $0$ or $1$. The interpretation in quantum mechanics is simple: either the particle is within the region (interval) $J_\cR$, in which case the measurement value $J_\cR=1$ will be found, or it is outside this region, and $J_\cR=0$ will be found. Particles are discrete objects - they are either inside or outside an interval. 

In our classical statistical Ising-type setting $J_\cR$ is a classical observable, given by 
\be\label{VV26}
J_\cR=\int_\cR N(x)=\sum_\cR N_L(x).
\ee
The sum $\sum_\cR$ extends over all lattice points within the region $\cR$. The classical observable $N_L(x)$ corresponds to a normalization of occupation numbers for the discrete lattice where $\big (N_{L,\epsilon}\big)_\tau=0,1$. The sum over species $N_L(x)=\sum\limits_\epsilon N_{L,\epsilon}(x)$, cf. eq. \eqref{FF8}, can therefore take in any classical state $\tau$ only the discrete values $\big(N_L(x)\big)_\tau=(0,1\dots, 8)$, according to the total number of eight species. In consequence, for any state $\tau$ of the classical statistical ensemble, $(J_\cR)_\tau$ is a positive integer or zero. According to the standard rule of classical statistics these integers describe the possible outcomes of measurements. 

For a one particle state the total particle number equals one,
\be\label{VV27}
1=\int\limits_V N(x)=\sum_VN_L(x),
\ee
where $\sum\limits_V$ extends now over all lattice points in the total volume. Since $\cR$ must be contained in $V$ the maximal allowed value for $J_\cR$ in a one-particle state is one, such that $(J_\cR)_\tau=0,1$ are the only possible values of the classical observable for such a state. This is precisely the quantum mechanical rule. No new postulate is necessary for this measurement in quantum mechanics - the quantum rule is inferred from the standard rule of classical statistics. Of course, the expectation value of $J_\cR$ for a one particle state is the same in the quantum mechanical and the classical statistical description 
\be\label{VV28}
\kl J_\cR\kr=\int_x\varphi^\dagger_D(x)J_\cR(x)\varphi_D(x)=
\int_\cR\varphi^\dagger_D(x)\varphi_D(x).
\ee

While our model connects the discrete particle aspects in quantum  mechanics directly to the discrete classical Ising-spins, the continuous wave aspects also arise in a natural way. The quantum wave function is continuous because probability distributions and associated classical wave functions are continuous (at least piecewise). As we have mentioned already, the characteristic interference effects for waves and the superposition arise from the linearity of the fundamental evolution equation in the classical wave function.

\section{Conclusions}
\label{Conclusions}
We have derived the Schr\"odinger equation for a quantum particle in a potential from a classical statistical ensemble for Ising-spins. The dynamics of the classical statistical system has to be specified by an  appropriate evolution equation for the probability distribution. For this purpose we employ the classical wave function which is defined as the positive or negative root of the probability distribution. The proposed evolution equation is a linear differential equation for the classical wave function. The wave function at time $t$ obtains from the wave function at $t'$ by a rotation - this guarantees the preservation of the normalization of the probability distribution. 

We have exploited a map between the classical wave function and a Grassmann wave function which is an element of a real Grassmann algebra. In turn, the time evolution of the Grassmann wave function can be associated to a Grassmann functional integral. This allows us to formulate the evolution equation for the classical wave function in terms of the action of a functional integral. The symmetries of the  model, as Lorentz symmetry and electromagnetic gauge symmetry for our model of Dirac spinors in an external electromagnetic field, can be easily implemented in this way. Since the classical wave function is real we have to formulate the model in terms of a real Grassmann algebra.

Our model is regularized on a lattice of space points. On the one hand, this guarantees that mathematical expressions are well defined for a finite number of lattice points, with continuum limit of an infinite number of points taken at the end. On the other hand, the concept of classical Ising-spins or associated occupation numbers at every  point $x$ is well defined. The discreteness of the particle aspects of quantum  mechanics can be traced back to the discrete occupation numbers that can only take the values zero or one. We also have used discrete time steps, and we have formulated the lattice action such that the Grassmann wave function $g(t+2\epsilon)$ obtains from $g(t)$ by a rotation. The time evolution is unitary not only in the limit $\epsilon\to 0$, but also for finite $\epsilon$. A unitary time evolution for infinitesimal time steps is easily achieved by any antisymmetric evolution generator $K_{\tau\rho}$ for the wave function,  $q_\tau(t)$, cf. eq. \eqref{P35}. Generating a unitary evolution also for finite smallest time steps imposes restrictions on the form of the lattice action. This, together with the requirement of a real Grassmann action, requires some care for the construction of the model and explains the specific form of the action \eqref{N1}-\eqref{B17} in comparison with other possible lattice actions.

The proposed evolution equation for the classical statistical ensemble of Ising-spins does not only lead to the Schr\"odinger equation for non-relativistic one-particle states. It entails the full dynamical equations for a quantum field theory of Dirac fermions in an external electromagnetic field. The dynamics of states with an arbitrary number of fermions, including the characteristic interference patterns for indistinguishable fermions in quantum mechanics, is correctly described.

At this point it seems worthwhile to ask some questions about the origin of characteristic features of quantum mechanics in our classical statistical setting. Particle-wave duality is realized by the discreteness of Ising-spins on one side, and the continuous probability distribution or classical wave function on the other side. Interference arises from the formulation of the basic evolution law in terms of the classical wave function. While the classical wave function is real, it can nevertheless take positive and negative values which can add to zero locally if two wave functions are added. The superposition principle or linearity of the quantum evolution finds a direct origin in the formulation of a dynamical law for the classical statistical ensemble that is linear in the classical wave function. The characteristic physics of phases in quantum mechanics is connected to the presence of a complex structure within the real Grassmann algebra. Planck's constant $\hbar$ appears purely as a conversion factor of units. The uncertainty relations can be obtained directly from the possible solutions of the Schr\"odinger equation.

Finally, one of the most characteristic elements of the mathematical formulation of quantum mechanics is the presence of a non-commutative product for operators and associated observables. These structures are obviously present in our formulation and clearly very useful for a discussion of solutions of the Schr\"odinger equation or Dirac equation. 
The essence of the emergence of non-commutative structures is the coarse graining of information \cite{3A}, \cite{CWP}, \cite{GR}.

This issue has not been addressed in the present note. It seems reasonable to expect, however, that a system that is governed by the Dirac equation on microphysical scales - say lattice distances shorter than the Planck length - will also show similar properties at ``macroscopic scales'' associated to coarse graining. (Such macroscopic scales can still be much smaller than all characteristic scales of atom physics or elementary particle physics.) The basic reason is that the form of the Dirac equation for one-particle states is essentially fixed by the symmetries. It will not be altered if the coarse graining respects the symmetries. While it remains an interesting task to perform this coarse graining explicitly, the main message of this note is already very clear at the present stage: Quantum field theory and quantum particles can be obtained from a classical statistical ensemble.


\section*{References}
\begin{thebibliography}{99}
\bibitem{DEC}H. D. Zeh, Found. Phys. {\bf 1} (1970) 69; W.~Zurek, Rev. Mod. Phys. {\bf 75} (2003) 715
\bibitem{GR}C. Wetterich, Annals of Phys. {\bf 325} (2010) 852; Ann. Phys. (Berlin) {\bf 522} (2010) 467; Journal of Phys. {\bf 174} (2009) 012008
\bibitem{CWP}C. Wetterich, Phys. Lett. {\bf A376} (2012) 706; Annals of Phys. {\bf 325} (2010) 1359; Ann. Phys. (Berlin) 522 (2010) 807; arXiv: 1003.0772; 
\bibitem{DL}Y. Couder, E.~Fort, Phys. Rev. Lett. {\bf 97} (2006) 154101; A.~Eddi, E.~Fort, F.~Moisy, Y.~Couder, Phys. Rev. Lett. {\bf 102} (2009) 240401
\bibitem{TH}G. t'Hooft, Int. J. Mod. Phys. {\bf A25} (2010) 4384
\bibitem{Bo}D. Bohm, Phys. Rev. {\bf 85} (1952) 166
\bibitem{Bell}J. S. Bell, Physica 1 (1964) 195
\bibitem{BS}J. Clauser, M. Horne, A. Shimony, R. Holt, Phys. Rev. Lett. {\bf 23} (1969) 880;~
J.~Bell, ``Foundations of Quantum Mechanics'', ed. B. d'Espagnat (New York: Academic, 1971) p. 171;~
J.~Clauser, M. Horne, Phys. Rev. {\bf D10} (1974) 526;~
J.~Clauser, A. Shimony, Rep. Prog. Phys. {\bf 41} (1978) 1881
%\bibitem{KS}S. Kochen, E. P. Specker, Journal of Mathematics and Mechanics {\bf 17} (1967), 59;\\
%N. D. Mermin, Phys. Rev. Lett. {\bf 65} (1990) 3373;\\
%A. Peres, J. Phys. A: Math. Gen. {\bf 24} (1991) L175;\\
%N. Straumann, arXiv: 0801.4931
\bibitem{15}C. Wetterich, arXiv: 1111.4115
\bibitem{CWF}C. Wetterich, Annals of Phys. {\bf 325} (2010) 2750; Annals of Phys. {\bf 326} (2011) 2243
\bibitem{CWES}C. Wetterich, arXiv: 1002.2593
\bibitem{Kop}B. Koopman, Proc. Nat. Acad. Sci. {\bf 17} (1931) 315; J. von Neumann, Ann. Math. {\bf 33} (1932) 587; {\bf 33} (1932) 789;~D.~Mauro, Phys. Lett. {\bf A315} (2003) 28
\bibitem{14}C. Wetterich, in ``Decoherence and Entropy in Complex Systems'', ed. T.~Elze, p. 180, Springer Verlag 2004, arXiv: quant-ph/0212031
\bibitem{CWMSab}C. Wetterich, Nucl. Phys. {\bf B852} [FS] (2011) 174
\bibitem{3A}C. Wetterich, arXiv:1005.3972

\end{thebibliography}
\end{document}